*Article*

# Correlation analysis technique of key parameters for transformer material inspection based on FP-tree and knowledge graph

Jing Xu[1] and Yongbo Zhang[1]

[1] Independent Researcher

**Abstract:** As one of the key equipment in the distribution system, the distribution transformer directly affects the reliability of the user power supply. The probability of accidents occurring in the operation of transformer equipment is high, so it has become a focus of material inspection in recent years. However, the large amount of raw data from sample testing is not being used effectively. Given the above problems, this paper aims to mine the relationship between the unqualified distribution transformer inspection items by using the association rule algorithm based on the distribution transformer inspection data collected from 2017 to 2021 and sorting out the key inspection items. At the same time, the unqualified judgment basis of the relevant items is given, and the internal relationship between the inspection items is clarified to a certain extent. Furthermore, based on material and equipment inspection reports, correlations between failed inspection items, and expert knowledge, the knowledge graph of material equipment inspection management is constructed in the graph database Neo4j. The experimental results show that the FP-Growth method performs significantly better than the Apriori method and can accurately assess the relationship between failed distribution transformer inspection items. Finally, the knowledge graph network is visualized to provide a systematic knowledge base for material inspection, which is convenient for knowledge query and management. This method can provide a scientific guidance program for operation and maintenance personnel to do equipment maintenance and also offers a reference for the state evaluation of other high-voltage equipment.

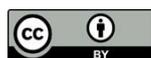

## 1. Introduction

High-quality power grid equipment materials are the primary prerequisite to ensure power grid safety. The quality inspection work of power grid materials plays an important role in the material quality supervision work for power grid companies. As a crucial control means of power material management, power grid material quality inspection connects the material production site and engineering construction site and is related to the safety and smooth operation of power grid engineering. It is of great significance to power construction, power grid security, and operation and maintenance.

Power grid material quality inspection refers to testing power equipment and materials supplied to the national power grid company through test witness, sampling, and third-party testing. With the development of the economy and society, the demand for electric power is increasing and the power grid is growing, which further leads to the expanding supply of materials needed for the laying of the power grid and the corresponding workload of material inspection is also growing year by year. Therefore, it is imperative to improve the efficiency of material inspection. The existing research mainly focuses on the condition maintenance, condition evaluation, and fault diagnosis of the equipment in the operation stage, and rarely involves the analysis and research of the quality of the equipment in the network stage. As a result, although a large number of inspection data can effectively characterize the performance of the equipment, it is limited to the judgment of equipment compliance and lacks effective analysis and application.

As one of the main equipment in the distribution system, the distribution transformer is directly related to the reliability of the user's power supply. There are many accidents in the operation, so it has become the focus of material inspection in recent years. Based on the unqualified inspection data of distribution transformer materials in recent years, the association rule algorithm is used to mine the association relationship between unqualified distribution transformer inspection items, sort out the key inspection items, and give the unqualified judgment basis of related items, thus clarifying the internal relationship between inspection items to a certain extent. It can improve the material inspection work's pertinence, effectiveness, and efficiency [1-3]. On this basis, based on the material equipment inspection report, the association relationship between non-conforming inspection items mined by association rules and expert knowledge, and using natural language processing technology from semi-structured data, the knowledge of entity, attribute, and relationship is extracted, and the knowledge graph of material equipment inspection management is constructed in Neo4j.

The power system is increasingly dependent on the technology of information. The data mining technology, which extracts hidden, unknown, non-trivial, and potential application value information or patterns from big data, is increasingly widely used in power systems, especially in typical big data scenarios [4-7]. Association rule mining can objectively analyze the hidden relationship between the influencing factors and the occurrence behavior, so it is more in line with the actual operation and production of the power system. Compared with other data mining methods, it is more widely used in the analysis of causality, such as the transformer fault diagnosis method based on association rules. Tan et al [8]. summarized and analyzed the influence variables of the transformer state, and studied the objective evaluation method of the transformer state based on association rule mining. Xie et al [9]. studied the application of the association rule mining method in power grid fault diagnosis. In addition, association rule mining has also been applied to the operation state analysis of equipment in power plants [10,11].

The knowledge graph is essentially a knowledge base or a knowledge network that connects and organizes entities and attributes through relationships. It identifies, filters, and deduces the complex relationships between things from unstructured knowledge, and efficiently completes the operation and analysis of relationships such as storage and query. It is widely used in search engines, social networks, medical education, power, and other fields. For the power grid, the knowledge graph can integrate the scattered knowledge within the power grid and effectively mine the useful potential rules in the large-scale text information in the power system. At the same time, the graph data structure of a knowledge graph can also provide great convenience for manual understanding. At present, the research of knowledge graph in the electric power field is still in its infancy, and the relevant literature mainly focuses on application exploration and macro framework design. Li et al. [12] sorted out the discrete scheduling knowledge according to the source code, configuration files, database, and expert experience of the dispatching automation system, and constructed the knowledge graph of the dispatching automation system, which is used for the dynamic search of business knowledge. Liu et al. [13] used an electric power dictionary to extract defect information entities and constructed a power equipment defect knowledge graph for fault defect information retrieval. Wang et al. [14] established an automatic generation system of security measures for intelligent substations by using a knowledge graph, which can improve the efficiency of secondary maintenance and the accuracy of safety measures to a certain extent. Wang et al. [15] extracted knowledge from power network dispatching rules, equipment failure emergency plans, and manual experience knowledge and constructed a power network fault dispatching knowledge graph to assist distribution network fault decision-making diagnosis. These studies provide relevant research ideas for this topic.

## 2. Materials and Methods

*2.1. Association rule algorithm technology*

2.1.1. Basic theory of association rules

Association rules are usually used to reflect the interdependence and correlation between one thing and other things. Association rules can be applied to a variety of scenarios. In the field of commercial sales, it can be used for cross-selling to get more income; in the medical field, possible treatment combinations can be found. All these belong to the problem of association rule mining. Here are a few basic concepts related to association rules.

The smallest indivisible unit of information (i.e. record) in a database is called an item (or item), represented by the symbol i, and the set of items is called an itemset. Let the set I = $\{i_1, i_2, ..., i_k\}$ be an itemset and the number of items in I is k. Then the set I is called a k-itemset. For example, the set {beer, diapers, milk powder} is a 3-itemset, and milk powder is an item. Each transaction is an itemset. Let I=$\{i_1, i_2, ..., i_k\}$ be the full set of all items in the database, and then the itemsets corresponding to each transaction ti is a subset of I. Transaction database T=$\{t_1, t_2, ..., t_n\}$ is a set consisting of a series of transactions with a unique identity. An association rule is an expression in the form of , where A and B are proper subsets of itemset I respectively, and, A is called the premise or leading condition, and B is called the result or successor of the association rule. The association rule reflects the law that when the item in A appears, the item in B also appears. Here, A and B do not refer to A single commodity, but to the itemset. For example, means that if a transformer fails to meet the quality requirements in two inspection items A and B, it will also fail in inspection item C.

The Support of an association rule is the ratio of the number of transactions in the transaction set that contain both items A and B to the total number of transactions in the transaction set. It reflects the probability that the items contained in A and B occur together in the transaction set and is calculated as follows.

$$\text{Support}(A \Rightarrow B) = P(A \cup B) = \frac{f(A \cup B)}{|D|} \times 100\% \quad (1)$$

The support reflects the validity of the association rule, which indicates the importance or the probability of occurrence of the association rule in the transaction database. The higher the support, the higher the association degree.

The Confidence of an association rule is the ratio of the number of transactions that contain both A and B to the number of transactions that contain A. Reflects the involvement in the affairs of A B of conditional probability P (B ∣ A), the calculation formula is:

$$\text{Confidence}(A \Rightarrow B) = P(B|A) = \frac{f(A \cup B)}{f(A)} \times 100\% \quad (2)$$

Confidence level reflects the certainty of association rules, said a credible degree of association rules, the higher the degree of confidence, the higher the credibility of the inference.

In general, the support and confidence values must be greater than or equal to artificial thresholds to indicate an association between items. If the support of an itemset is greater than or equal to the minimum support, it is called a frequent itemset. A frequent itemset is a collection of items that often appear together, implying that certain items always occur in pairs or groups. Finding frequent itemsets is the first step to finding strong association rules. If the confidence of frequent itemsets is greater than or equal to the minimum confidence, it would be called a strong association rule. Only strong association rules have practical significance, so the association rules usually refer to strong association rules.

2.1.2. FP-growth algorithm

The purpose of mining association rules is to find all associated attributes related to an event by taking an event A as the entry point, to maximize the acquisition of attribute connections, that is, strong association rules. Usually, the validity of the rule is verified according to the lifting degree. The algorithm includes the following two main steps: finding frequent itemsets and derivation of strong association rules. Common association rule algorithms include Apriori and FP-growth.

The FP-growth algorithm does not need to generate candidate itemsets and only needs to scan the dataset twice. The first time scans the dataset to get frequent 1-itemsets. In the second scan of the data set, the frequent 1- itemsets are used to filter the infrequent items in the data set, and the FP-Tree is generated (as shown in Figure 1), and then the recursive algorithm is executed on the FP-Tree to mine all the frequent itemsets. Experimental results show that the FP-Growth algorithm is faster than the Apriori algorithm by an order of magnitude.

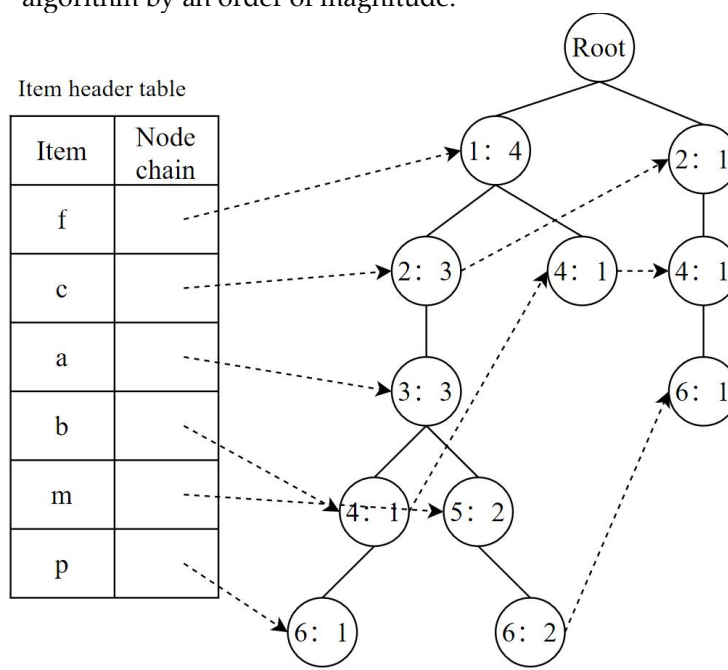

**Figure 1.** FP-tree structure diagram

FP-Growth algorithm (as shown in Figure 1) is a typical pattern growth algorithm, it is also the first pattern growth algorithm, and many subsequent pattern growth algorithms are based on it. Compared with the Apriori algorithm, the FP-Growth algorithm is more efficient in many cases (such as a low user-defined minimum threshold, a large number of items in the dataset, or a large number of long transaction items in the dataset).

*2.2. Construction of knowledge graph for inspection and management of materials and equipment*

Knowledge graph is a technical method that uses graph models to describe the relationship between knowledge and things. The knowledge graph is composed of nodes and edges. Nodes can be entities, such as a device, a person, or abstract concepts, such as the knowledge graph, artificial intelligence, and so on. Edge is generally the relationship between entities, such as teachers students, and family members. It can also be the attributes of the entity, such as device name, device type, etc.

Knowledge graph is generally expressed in the form of a triple, namely (entity A, relation, entity B) or (entity, relation, attribute). Entities and attributes are nodes, and the relationship is to connect the directed edges of two nodes. Triples are connected by entities or attributes and their relationships to effectively organize large-scale network information with minimum cost. The knowledge graph method realizes the visual storage, organization, and management of entities, attributes, and relationships, and forms a structured knowledge representation. This kind of graph data structure is easy to understand and accept by people. It is also easy to be recognized and processed by computers.

The construction methods of a knowledge graph can usually be divided into three modes: top-down, bottom-up, and hybrid. The knowledge graph of inspection and management of electric equipment belongs to the domain knowledge graph, which is usually constructed by the combination of top-down and bottom-up. That is, first define the ontology layer, then extract knowledge in the data layer, update the ontology layer repeatedly, and finally add the new knowledge to the ontology to complete the graph construction. Based on the material equipment inspection corpus, this paper constructs the knowledge graph of material equipment inspection management and realizes the automatic matching of material inspection data. The specific process is shown in Figure 2.

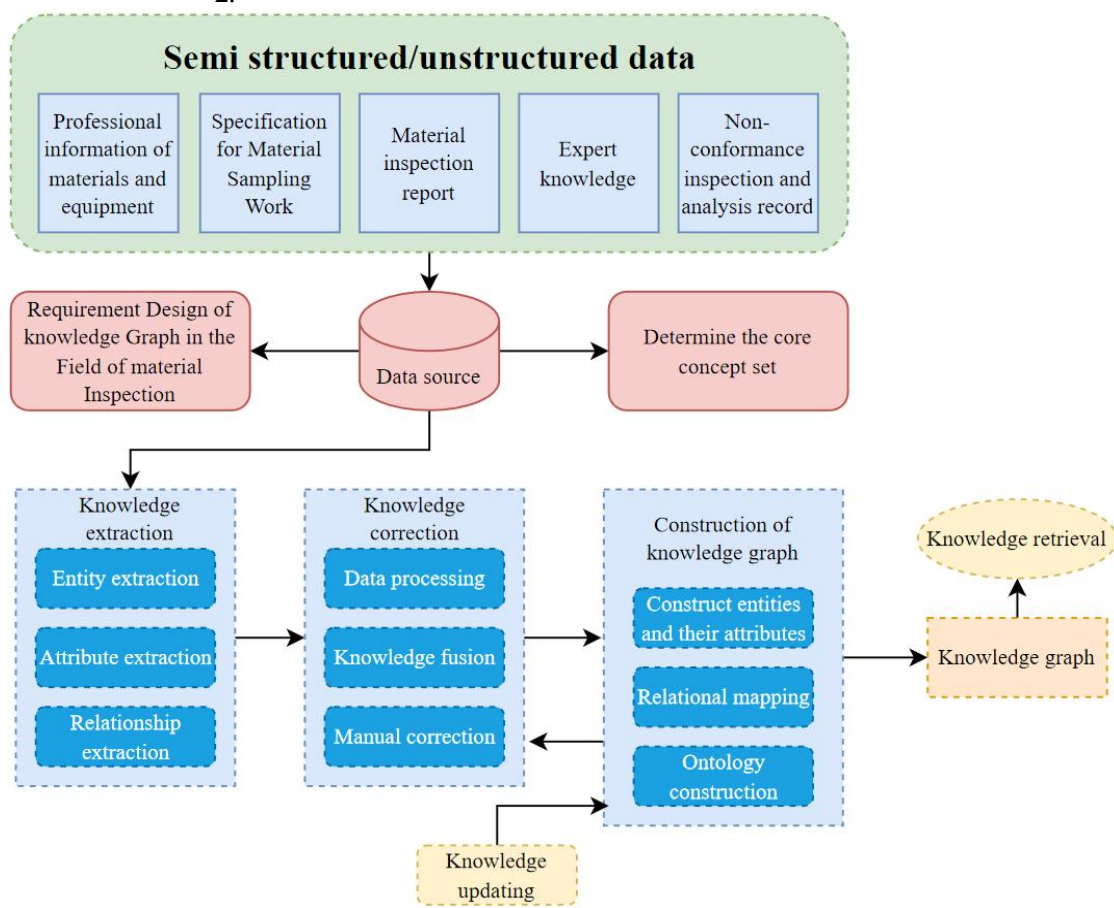

**Figure 2.** Construction process of knowledge graph for material and equipment inspection and management

After defining the design requirements of the knowledge graph of material and equipment inspection and management, through the research and analysis of the relevant knowledge in the field of material and equipment inspection, the important terms and concepts are summarized. Meanwhile, the unstructured and semi-structured data such as material inspection reports are summarized. On the premise of defining the rationality of concept division, this paper accurately describes the knowledge system of material inspection and the relationship between knowledge points and corrects ambiguous and confusing terms. Finally, an accurate and perfect core concept set of material inspection is formed.

Then we need to extract the entity and its attributes from the data. The purpose of entity extraction is to extract the words that represent the entity and the attributes and attribute values of the entity from the corpus of material inspection and analysis. It needs to be tagged in part of speech in this process. Through the above process, you can use triples (entities, attributes, attribute values) to describe the attributes of entities and their attribute values. The role of attributes in a knowledge graph is to describe entities. However, entity attribute filling is faced with the following dilemmas:

1) Because most of the entities in the quality inspection of materials and equipment are power equipment or devices, a large number of related attributes and indicators will be collected in the inspection. If all these data are added to the knowledge graph, the scale of the graph will be too large and the upper application will be affected.

2) The collected data have some incompleteness, and there may be a lack of indicators, so what kind of indicators as attributes need to be further studied, and the attribute values of the inspection items are not extracted in this paper.

The entity types related to the knowledge graph of material equipment inspection and management are shown in Table 1. Among them, the type and cause of the problem are based on the inspection report of all kinds of materials from 2017 to 2021, which is matched and extracted by the electric power professional dictionary, and finally collated and summarized with expert knowledge. The related inspection items are the results of the association relationship among the inspection items obtained after the analysis of association rules.

**Table 1.** Entity types of Knowledge Graph of material equipment Inspection and Management

| Entity type | Example |
| --- | --- |
| Equipment | Distribution transformers, Power cables, Overhead insulated wires, etc. |
| Component | Iron core, insulating sleeve, Cooling device, Tap-changer, etc. |
| Inspection index | The dielectric loss factor exceeds the standard, The resistance imbalance rate exceeds the standard, The no-load loss exceeds the standard. |
| Inspection items | Insulating oil test, winding resistance measurement, short circuit impedance and load loss measurement, etc. |
| Cause of defect | Impurities in oil, Bumps lead to oil leakage, Unqualified gaskets, etc. |
| Type of defect | Raw material problems, Assembly process problems, Process control problems, etc. |
| Related inspection items | Insulating oil test, Wind resistance measurement, Short circuit impedance and load loss measurement, etc. |

Then we need to extract the semantic relationship between entities from the corpus of material inspection and analysis and explain the relationship between entities. This is an essential step in providing query functions and visualization [16]. The above process is usually called relationship extraction, and the relationship is generally expressed in the form of a triple, such as (entity 1, relationship, entity 2) or (entity, relation, attribute). Entities and attributes are nodes, and the relationship is to connect the directed edges of two nodes. Triples are connected by common entities or attributes and their relationships to form a knowledge graph with a network structure. The triple is connected by common entities or attributes and their relationships to form a knowledge graph with a network structure. The triplet relationship in the knowledge graph of material equipment inspection and management, taking the distribution transformer as an example, is shown in Table 2.

Finally, the knowledge graph of material and equipment inspection is constructed and managed by graph nodes and relational edges based on the Neo4j graph database. Neo4j is a high-performance NoSQL graph database, which can store structured data on the topology graph and realize the centralized management of a large number of data. The graph model is composed of nodes and relations, nodes store entity information, each node is linked by the relationship between them, and the attributes and labels of nodes and relationships are stored in the form of key-value pairs.

**Table 2.** The Triple relationship of each Test item of the Distribution Transformer

| Serial Number | Triple relation |
| --- | --- |
| 1 | (Distribution transformers, Components, Iron cores) |
| 2 | (Iron core, Inspection items, No-load loss and no-load current measurement) |
| 3 | (No-load loss and no-load current measurement, Inspection indicators, no-load loss exceeding standards) |
| 4 | (Excessive no-load loss, Cause of defect, Margin too small) |
| 5 | (Margin too small, Defect type, Design structure) |
| 6 | (External withstand voltage, Related items, Lightning impulse) |

**3. Results and Discussion**

*3.1. Correlation analysis of inspection items based on association rules*

**Table 3.** Test result data of distribution transformer

| Device number | Nonconforming inspection item |
| --- | --- |
| WB2022060389 | Short circuit impedance and load loss measurement |
| WB2022060386 | Sound level determination |
| WB2022060387 | External voltage, lightning impact, short circuit, bearing capacity |
| WB2022060389 | Temperature rise, sound level determination, short circuit bearing capacity |
| WB2022060342 | Short circuit tolerance |
| WB2022060335 | Short circuit tolerance |
| WB2022060345 | No-load loss and no-load current measurement, short-circuit bearing capacity |
| WB2022060365 | Sound level determination, short circuit bearing capacity |
| WB2022050471 | Short circuit impedance and load loss measurement |
| WB2022050579 | Winding resistance measurement, short circuit bearing capacity |
| WB2022050457 | No-load loss and no-load current measurement, short-circuit impedance and load loss measurement |
| …… | …… |

In this paper, based on the distribution transformer detection data collected from 2017 to 2021, the association rules algorithm is used to mine the association relationship between unqualified inspection items. To a certain extent, this can clarify the internal relationship between the inspection items, to improve the efficiency of material inspection.

**Table 4.** Results of distribution transformer association rules

| Rule lead term | Rule follow-ups | Support | Confidence |
|---|---|---|---|
| External pressure test | Lightning impact test | 35.5% | 95.6% |
| Induction withstand pressure test | Lightning impact test | 10.21% | 95.21% |
| Insulating oil test | External pressure test | 27.34% | 95.14% |
| Temperature rise test | Protection class measurement | 15.25% | 92.32% |
| Winding resistance measurement | No-load loss and no-load current measurement | 35.25% | 92.18% |
| Voltage ratio measurement | Induction voltage test | 31.51% | 90.87% |
| Pressure seal test | Insulating oil test | 27.83% | 90.34% |
| Insulation resistance measurement | Load loss and short-circuit impedance measurements | 24.26% | 90.27% |
| External pressure test | Insulating oil test | 17.73% | 89.92% |
| Temperature rise test | Short circuit tolerance | 32.72% | 89.38% |
| Winding resistance measurement | Load loss and short-circuit impedance measurement | 24.71% | 89.12% |
| Winding resistance measurement | Induction voltage test | 16.78% | 88.76% |
| Voltage ratio measurement | Winding resistance measurement | 12.23% | 88.68% |
| Winding resistance measurement | Short circuit tolerance | 27.42% | 88.25% |
| Induction withstand pressure test | Insulation resistance measurement | 11.32% | 87.98% |

According to the requirements of State Grid Corporation of China's sampling work for distribution transformer voltage, the main inspection items for distribution transformer connection to the grid include: (1) insulation oil test; (2) insulation resistance measurement; (3) winding resistance measurement; (4) voltage ratio measurement; (5) no-load loss and no-load current measurement; (6) load loss and short-circuit impedance measurement; (7) high voltage insulation test (including external voltage test, induction voltage test, lightning impact test); (8) Pressure seal test; (9) temperature rise test; (10) sound level measurement; (11) short-circuit bearing capacity. Its analysis process is as follows.

**Step 1:** Input the data set of unqualified inspecting items over the years, and set the minimum support and minimum confidence thresholds. The data set includes the names of materials and equipment, equipment serial numbers, and unqualified inspection items. The transaction dataset composed of N transaction sets D={$T_1$, $T_2$, ..., $T_N$}, where $T_j$ (j = 1, 2, ..., N) is the j-th transaction set of the transaction data set D. The elements forming the transaction set $T_j$ are called the item $I_k$ (k= 1, 2, ..., p) of the transaction set, I={$I_1$, $I_2$, ..., $I_p$} is a set composed of all different items in dataset D. The item Ik in the transaction set is the name of the inspection item, and all the unqualified inspection items of the same equipment form a transaction set $T_j$. The specific data form is shown in Table 3.

**Step 2:** The minimum support threshold was set to 0.01, and the minimum confidence was 0.6. The Apriori algorithm was used to mine the association relationship between unqualified inspection items, and the running time was 0.68s. The running time of the FP-Growth algorithm is 0.08s, and 15 strong association rules satisfying the minimum support and minimum confidence thresholds are obtained, and the results are shown in Table 4 (theoretically, the results of the two algorithms should be the same).

**Step 3:** Explain the meaning of the rule, take {external pressure} → {lightning impact} as an example, it appears 53 times together, the support is 35.5%, and the confidence is 95.6%, indicating that when the external pressure is unqualified. It can be inferred that lightning impact may also be unqualified, and the credibility of the inference is 95.6%. Through such association inference and credibility value, the judgment basis of unqualified related items can be given, to delete the inspection items and improve the efficiency of inspection. In addition, the above results can be used as one of the data sources of the material equipment monitoring management knowledge graph to construct the network relationship graph.

## *3.2. Visualization of knowledge graph*

According to the collected material inspection and analysis corpus, through entity, relationship, and other knowledge extraction and knowledge storage, this paper achieves the construction of a knowledge graph. In this paper, taking the relevant detection information of the distribution transformer as the data source, an example of visualization of the distribution transformer graph is given as shown in Figure 3. The nodes in the diagram are composed of the components of the distribution transformer, inspection items, testing indicators, causes of problems, types of problems, and so on. Different types of nodes are distinguished by different colors when they are visualized. Through the query language, the Neo4j diagram database can clearly show the related entities and their relationships of distribution transformer inspection, which is helpful for knowledge query and question diagnosis.

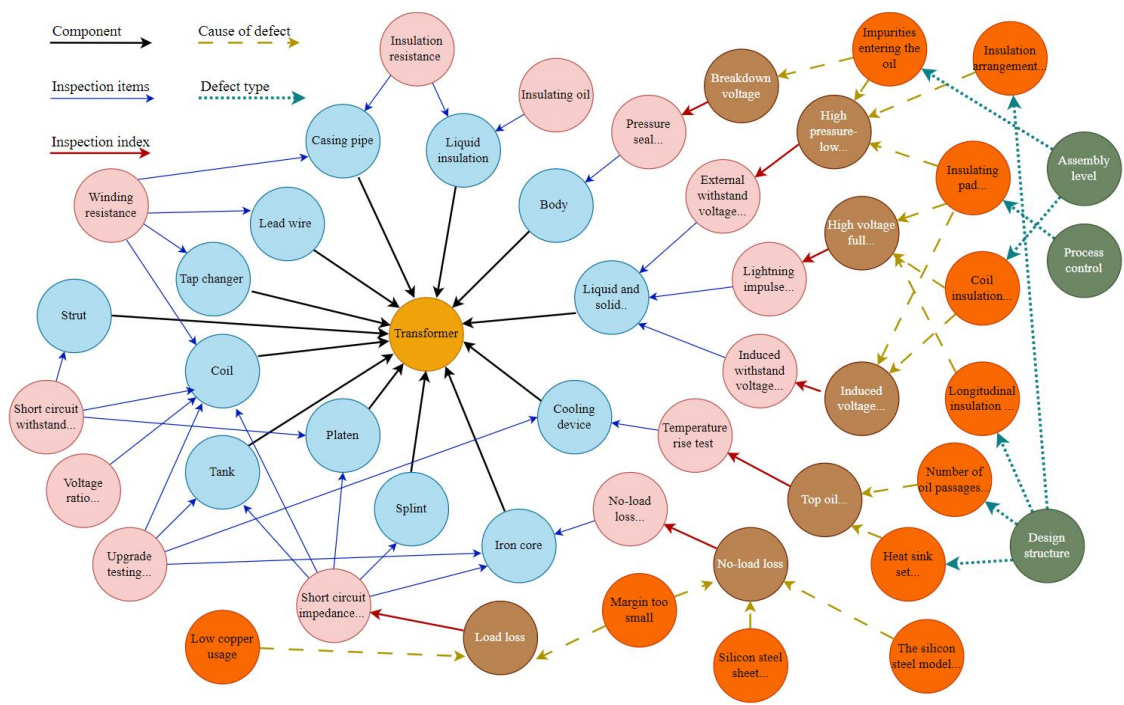

**Figure 3.** An example of Visualization of knowledge Graph for material equipment Inspection Management

## 4. Conclusions

Material inspection is related to the safety and stable operation of power grid engineering. Transformer equipment has a high probability of accidents in operation, so it has become the focus of material inspection. However, a large number of original record information of key parameters based on sampling operation specifications have not been effectively used. Given the above problems, this paper excavates the relationship between unqualified inspection items by analyzing the material inspection data of Shanxi Electric Power Company over the years. In addition, based on the knowledge graph technology, the knowledge graph of material and equipment inspection and management is constructed. By analyzing association rules, a knowledge graph of equipment detection management for each unqualified detection item is constructed. Through the analysis of the association rules of each unqualified detection project and the construction of a knowledge graph, we can comprehensively improve the carrying capacity of the power grid material inspection system, and effectively solve the problem of insufficient capacity in work efficiency, in-depth analysis and so on.

In this paper, the advantages and disadvantages of the Apriori algorithm and FP-Growth algorithm in the application of this problem are compared through experiments. It can be seen from the experimental results that the FP-Growth algorithm is significantly better than the Apriori algorithm for this problem. With the method proposed in this paper, the advanced nature and rationality of material inspection can be significantly improved, and the role of power grid material quality inspection as an escort for power construction, operation, and maintenance can be brought into full play.

**Funding:** This research received no external funding.

**Institutional Review Board Statement:** Not applicable.

**Informed Consent Statement:** Not applicable.

**Conflicts of Interest:** The authors declare no conflicts of interest.